\documentstyle[prl,aps,epsf,floats]{revtex}
\begin{document}
\draft
\twocolumn[\hsize\textwidth\columnwidth\hsize\csname @twocolumnfalse\endcsname
\title{Photoemission spectroscopy and sum rules in dilute 
electron-phonon systems} 
\author{P.\,E.\,Kornilovitch}
\address{
Hewlett-Packard Laboratories, MS 1L-12, 1501 Page Mill Road, Palo Alto, 
California, 94304
}

\date{\today}
\maketitle

\begin{abstract}

A family of exact sum rules for the one-polaron spectral function in the
low-density limit is derived.  An algorithm to calculate energy 
moments of arbitrary order of the spectral function is presented.  
Explicit expressions are given for the first two moments of a model 
with general electron-phonon interaction, and for the first four 
moments of the Holstein polaron.  The sum rules are linked to 
experiments on momentum-resolved photoemission spectroscopy.  The 
bare electronic dispersion and the electron-phonon coupling constant 
can be extracted from the first and second moments of spectrum.  
The sum rules could serve as constraints in analytical and 
numerical studies of electron-phonon models.

\end{abstract}
\vskip2pc]
\narrowtext

Statistical analysis of spectra is an important branch of
photoemission spectroscopy.  It has been known that the energy 
moments of the spectral density of atoms can be expressed via 
the atomic quantum numbers \cite{Fano,Thole}.  Such integral 
relations, or sum rules, provide direct link between experimental 
data and the internal state of the atoms.  Similarly, in 
crystals analogous sum rules yield valuable information about 
the Bloch states.  This is particularly true in case of 
two-dimensional systems where the in-plane momentum of the 
electron does not change in the photoemission process.  
It allows the spectrum to be recorded for each in-plane 
momentum individually.  In recent years, the angle-resolved 
photoemission spectroscopy became a powerful experimental 
technique to study the electronic structure of quasi-two-dimensional 
compounds such as the high-temperature superconductors and layered 
colossal magnetoresistive oxides \cite{Lanzara,Dessau,Kaminski}.  

This paper discusses statistical analysis of spectra from systems 
with strong electron-phonon (el-ph) interaction.  In such systems, 
the mobile carriers deform the lattice and form polarons which are 
bound states of the carriers and the lattice deformations.  The 
polarons then define the low-energy physics of the system, in 
particular the equilibrium chemical potential.  In inverse 
photoemission (IPES), an electron with momentum {\bf k} is injected 
into the system from outside.  Initially it feels the bare electronic 
states in the undeformed lattice and only then begins to deform the 
lattice around itself.  Thus it should be expected that the energy 
moments of the IPES intensity could provide information about the 
initial stages of the deformation process and therefore about the 
strength of the interaction.  It turns out that for low density of 
polarons the moments of IPES intensity can be calculated analytically 
for a wide class of model el-ph interactions.  It leads to a series 
of exact sum rules which the spectral density, experimental or 
theoretical, must satisfy.  Below I derive the sum rules and discuss 
their possible applications.

Consider a single band, one optical branch Hamiltonian with an 
el-ph interaction of arbitrary profile \cite{Kornilovitch,AleKor}
\begin{eqnarray}
{\cal H} = \sum_{\bf k} ( \varepsilon_{\bf k} - \mu )
c^{\dagger}_{\bf k} c_{\bf k} & - & \sqrt{2} \Omega \sum_{\bf n m} 
g({\bf n} - {\bf m}) c^{\dagger}_{\bf n} c_{\bf n} x_{\bf m}
  \nonumber   \\
& + & \frac{\Omega}{2} \sum_{\bf m} ( -\partial^2_{\bf m} + x^2_{\bf m} ).
\label{sone}
\end{eqnarray}
Here $\mu$ is the chemical potential, 
$x_{\bf m} = \xi_{\bf m} \sqrt{\frac{M\Omega}{\hbar}}$ is the 
dimensionless displacement of an oscillator at {\bf m}-th lattice 
site, $M$ and $\Omega$ are the mass and the frequency of the oscillators,
respectively, $\partial_{\bf m} \equiv \partial/\partial x_{\bf m}$, 
$c_{\bf n} = N^{-1/2} \sum_{\bf k} e^{i {\bf k} {\bf n}} c_{\bf k}$,
and $N$ is the number of lattice sites.  $g({\bf n} - {\bf m})$ is
the dimensionless coupling constant proportional to the force with
which an electron at site {\bf n} acts on oscillator {\bf m}.  For 
simplicity, the fermion spin and multiple phonon polarizations are not 
included explicitly in Eq.(\ref{sone}).  They are not essential for the 
purposes of the paper and could be added if necessary.  The canonical
Holstein model \cite{Holstein} follows from the above Hamiltonian 
with the choice $g({\bf n} - {\bf m}) = g_0 \delta_{\bf nm}$. 

Within the sudden approximation, direct and inverse photoemission 
intensities are proportional to the fermion spectral functions 
$A_{-}({\bf k},\omega)$ and $A_{+}({\bf k},\omega)$, respectively 
(hereafter notation $\hbar = 1$ is used):
\begin{equation}
A_{-}({\bf k},\omega) = \frac{2\pi}{Z} \!
\sum_{ij} e^{-\beta {\cal E}_i}  
\vert \langle j | c_{\bf k} | i \rangle \vert^2 
\delta( \omega + {\cal E}_j - {\cal E}_i ) ,
\label{stwo}
\end{equation}
\begin{equation}
A_{+}({\bf k},\omega) = \frac{2\pi}{Z} \! 
\sum_{ij} e^{-\beta {\cal E}_i}  
\vert \langle j | c^{\dagger}_{\bf k} | i \rangle \vert^2 
\delta( \omega + {\cal E}_i - {\cal E}_j ) .
\label{sthree}
\end{equation}
Here $\omega$ is the energy relative to the chemical potential,
$| i \rangle$ and $| j \rangle$ are exact eigenstates of 
${\cal H}$, with energies ${\cal E}_i$ and ${\cal E}_j$, 
$Z = \sum_n e^{-\beta {\cal E}_i}$ is the grand canonical 
partition function, and $\beta = (k_B T)^{-1}$.  The validity 
of the sudden approximation as well as the additional electron-photon
matrix element, the Coulomb effects and other complications,
are not going to be discussed in this paper, see e.g. \cite{Joynt}.
It follows from Eqs.~(\ref{stwo}) and (\ref{sthree}) that the 
two spectral functions obey the fundamental relation
\begin{equation}
A_{+}({\bf k},\omega) = e^{\beta \omega} A_{-}({\bf k},\omega) .
\label{sfour}
\end{equation}
Thus, $A_{+}$ can be found if $A_{-}$ is known with sufficient
accuracy, and visa versa.  In terms of the symmetrized function
$A = A_{+} + A_{-}$ they are expressed as 
$A_{-}({\bf k},\omega) = (e^{\beta \omega} + 1)^{-1} A({\bf k},\omega)$ 
and $A_{+}({\bf k},\omega) = 
(e^{- \beta \omega} + 1)^{-1} A({\bf k},\omega)$.

Several theoretical results for the polaron spectral function are 
available.  The on-site version of the Holstein model 
($\varepsilon_{\bf k} \equiv 0$) is exactly solvable, see
e.g. Ref.~\cite{Mahan}.  For the Holstein model with
non-zero kinetic energy the spectral function was calculated
by Alexandrov and Ranninger in the Lang-Firsov approximation 
\cite{AleRan}.  Later their result was generalized to the 
long-range interaction (\ref{sone}) by Alexandrov and Sricheewin
\cite{Sricheewin}.  For the more complex Jahn-Teller interaction
the polaron spectral function was given by Perebeinos and Allen
\cite{Perebeinos}.  The spectral functions mentioned are
approximate and all have the form of a sequence of delta-functions.  

The formalism of statistical moments is best introduced via the 
time-dependent correlation functions $K_{\pm}({\bf k},t)$ that are 
Fourier transforms of $A_{\pm}({\bf k},\omega)$:
\begin{eqnarray}
K_{+}({\bf k},t) & = & \int^{\infty}_{-\infty} 
\frac{d\omega}{2\pi} e^{-i\omega t} A_{+}({\bf k},\omega) \nonumber \\
                 & = & \frac{1}{Z} \sum_j e^{-\beta {\cal E}_j}
\langle j | e^{i {\cal H} t} c_{\bf k} e^{-i {\cal H} t} 
c^{\dagger}_{\bf k} | j \rangle  \nonumber \\  
& \equiv & \langle e^{i {\cal H} t} c_{\bf k} e^{-i {\cal H} t}
c^{\dagger}_{\bf k} \rangle ,
\label{sseven}
\end{eqnarray}
\begin{eqnarray}
K_{-}({\bf k},t) & = & \int^{\infty}_{-\infty} 
\frac{d\omega}{2\pi} e^{-i\omega t} A_{-}({\bf k},\omega) \nonumber \\
                 & = & \frac{1}{Z} \sum_j e^{-\beta {\cal E}_j}
\langle j | c^{\dagger}_{\bf k} e^{i {\cal H} t} c_{\bf k} 
e^{-i {\cal H} t} | j \rangle   \nonumber \\
& = & \langle c^{\dagger}_{\bf k} e^{i {\cal H} t} 
c_{\bf k} e^{-i {\cal H} t} \rangle .
\label{seight}
\end{eqnarray}
Taking the $m$-th derivative with respect to time $t$ and setting $t=0$  
one obtains
\begin{eqnarray}
{\cal M}^{+}_m({\bf k}) & \equiv & 
\int^{\infty}_{-\infty} \frac{d\omega}{2\pi} 
\omega^m A_{+}({\bf k},\omega) \nonumber \\
& = & \langle  
\underbrace{\left[ \left[ \left[ c_{\bf k} , {\cal H} \right] ,  
{\cal H} \right] \ldots , {\cal H} \right] }_{m \;{\rm times}}
c^{\dagger}_{\bf k} \rangle ,
\label{snine}
\end{eqnarray}
\begin{eqnarray}
{\cal M}^{-}_m({\bf k}) & \equiv & 
\int^{\infty}_{-\infty} \frac{d\omega}{2\pi} \omega^m 
A_{-}({\bf k},\omega) \nonumber \\
& = & \langle   
c^{\dagger}_{\bf k} \underbrace{\left[ \left[ \left[ c_{\bf k} ,
{\cal H} \right] , {\cal H} \right] \ldots , {\cal H} \right] 
}_{m \;{\rm times}} \rangle .
\label{sten}
\end{eqnarray}
The moments of the symmetrized function $A({\bf k},\omega)$ are
defined as ${\cal M}_m({\bf k}) = {\cal M}^{+}_m({\bf k}) +
{\cal M}^{-}_m({\bf k})$.

The simplest sum rule is the normalization property of $A({\bf k},\omega)$ 
which is common to any one-particle fermionic spectral function
\begin{equation}
{\cal M}_0({\bf k}) = \langle c_{\bf k} c^{\dagger}_{\bf k} + 
c^{\dagger}_{\bf k} c_{\bf k} \rangle = 1. 
\label{seleven}
\end{equation}
The first moment of $A({\bf k},\omega)$ is given by
\begin{eqnarray}
[c_{\bf k}, {\cal H}] & = & (\varepsilon_{\bf k} - \mu ) c_{\bf k} 
\nonumber \\
& - & \frac{\sqrt{2} \Omega}{N} \sum_{\bf n m k'}
e^{i({\bf k}'-{\bf k}){\bf n}} g({\bf n} - {\bf m}) 
x_{\bf n} c_{\bf k'} .
\label{stwelve}
\end{eqnarray}
\begin{eqnarray}
{\cal M}_1({\bf k}) & = &  
\langle [c_{\bf k}, {\cal H}] c^{\dagger}_{\bf k} + 
c^{\dagger}_{\bf k} [c_{\bf k}, {\cal H}] \rangle  \nonumber \\
& = & (\varepsilon_{\bf k} - \mu ) - \frac{\sqrt{2} \Omega}{N} 
\sum_{\bf n m} g({\bf n} - {\bf m}) \langle x_{\bf m} \rangle  . 
\label{sthirteen}
\end{eqnarray}
The mean displacement of the oscillators $\langle x_{\bf m} \rangle$
depends only on the total number of carriers in the systems 
$\langle n \rangle$, and can be calculated as follows.  Write
the displacement operator as 
$x_{\bf m} = \langle x_{\bf m} \rangle + y_{\bf m}$ and 
substitute in ${\cal H}$.  Then minimization of the free energy
$F = - \beta^{-1} \ln \langle e^{-\beta {\cal H}} \rangle$
with respect to $\langle x_{\bf m} \rangle$ yields  
\begin{eqnarray}
\langle x_{\bf m} \rangle & = & - \langle y_{\bf m} \rangle 
+ \sqrt{2} \sum_{\bf n} g({\bf n} - {\bf m}) 
\langle  c^{\dagger}_{\bf n} c_{\bf n} \rangle  \nonumber \\
& = & \sqrt{2} \langle n \rangle \sum_{\bf n} g({\bf n}) ,
\label{sfourteen}
\end{eqnarray}
because $\langle y_{\bf m} \rangle = 0$ by definition, and the mean 
density $\langle c^{\dagger}_{\bf n} c_{\bf n} \rangle = 
\langle n \rangle$ does not depend on coordinate ${\bf n}$.  
Substitution in Eq.~(\ref{sthirteen}) results in: 
\begin{equation}
{\cal M}_1({\bf k}) = (\varepsilon_{\bf k} - \mu ) - 
2 \Omega \langle n \rangle 
\left[ \sum_{\bf n} g({\bf n}) \right]^2 .
\label{sfifteen}
\end{equation}
This is an exact relation for the one-particle spectral function.
It is a generalization of the sum rule derived by Mahan for the 
on-site polaron \cite{Mahan}, and by Perebeinos and Allen for the
Jahn-Teller polaron \cite{Perebeinos}, to non-zero polaron density.  
The main feature of these results is that the mean value of the spectral
function is not affected by interaction with phonons.  This 
conclusion is non-trivial in view of significant changes of the
spectral function itself, especially in the strong coupling regime.
As follows from Eq.~(\ref{sfifteen}), in the many-body case the 
first moment ${\cal M}_1({\bf k})$ acquires a correction that is linear 
in polaron density.  Proceeding with the calculation one obtains 
for the second moment
\begin{eqnarray}
[ [ c_{\bf k}, {\cal H}] , {\cal H} ] & = &  
(\varepsilon^2_{\bf k} - \mu ) c_{\bf k}           \nonumber \\                
& - & \frac{\sqrt{2} \Omega}{N} \sum_{\bf k' n m}
e^{i({\bf k' - k}){\bf n}} g({\bf n} - {\bf m})    \nonumber \\
& \times & [(\varepsilon_{\bf k} + \varepsilon_{\bf k'} - 2\mu) 
x_{\bf m} + \Omega \partial_{\bf m} ] c_{\bf k'}   \nonumber \\
& + & \frac{2 \Omega^2}{N} \sum_{\bf k' n m m'}
e^{i({\bf k' - k}){\bf n}} g({\bf n} - {\bf m})    \nonumber \\
& \times & g({\bf n} - {\bf m'}) x_{\bf m} x_{\bf m'} c_{\bf k'} , 
\label{ssixteen}
\end{eqnarray}
\begin{eqnarray}
{\cal M}_2({\bf k}) & = & 
\langle [ [ c_{\bf k}, {\cal H}] , {\cal H} ] c^{\dagger}_{\bf k} + 
c^{\dagger}_{\bf k} [ [ c_{\bf k}, {\cal H}] , {\cal H} ] \rangle 
= (\varepsilon_{\bf k} - \mu )^2 \nonumber \\
& - & \frac{\sqrt{2} \Omega}{N} 
\sum_{\bf n m} g({\bf n} - {\bf m}) 
[ 2( \varepsilon_{\bf k} - \mu ) \langle x_{\bf m} \rangle +
\Omega \langle \partial_{\bf m} \rangle          \nonumber \\
& + & \frac{2\Omega^2}{N} \sum_{\bf n m m'}
g({\bf n} - {\bf m}) g({\bf n} - {\bf m'})
\langle x_{\bf m} x_{\bf m'} \rangle . 
\label{sseventeen}
\end{eqnarray}
Unfortunately, the mean values $\langle x_{\bf m} x_{\bf m'} \rangle$ 
cannot be calculated analytically which limits application of the last 
expression.  In principle, such averages could be computed numerically 
by various techniques, rendering Eq.~(\ref{sseventeen}) a true sum 
rule. (It is worth noting in this respect that in the case of the Hubbard 
model, the first and second moments of the spectral function were 
calculated analytically by Nolting \cite{Nolting}.  This finding enabled 
him to develop the moment approach to the Hubbard model.)

Further progress can be made by going to the low density limit 
$\langle n \rangle \ll 1$.  It follows from definitions (\ref{stwo})
and (\ref{sthree}) that $A_{-} \propto \langle n \rangle$ while
$A_{+} \propto 1 - \langle n \rangle$.  Therefore 
${\cal M}^{-}_m \ll {\cal M}^{+}_m$ so that ${\cal M}^{-}_m$ could be
dropped from Eqs.~(\ref{seleven}), (\ref{sthirteen}), and (\ref{sseventeen}).
More importantly, all the phonon operators can now be averaged easily.
Indeed, in the low-density limit the lattice is largely undeformed.
The function $A_{+}$ describes the process when an electron is created
in an {\em empty} lattice and removed later.  Thus whatever phonon 
operators appear as a result of multiple commutation with the Hamiltonian, 
they have to be averaged over the system of undeformed {\em independent} 
harmonic oscillators, which is always possible.  For instance, the mean 
values entering Eq.~(\ref{sseventeen}) are 
$\langle x_{\bf m} \rangle = \langle \partial_{\bf m} \rangle = 0$,  
$\langle x_{\bf m} x_{\bf m'} \rangle = \frac{1}{2} 
\coth (\frac{1}{2}\beta\Omega) \delta_{\bf mm'} $.  Collecting all 
the results one obtains the following sum rules for the spectral
function $A_{+}({\bf k},\omega)$:
\begin{eqnarray}
{\cal M}^{+}_0({\bf k}) & = & 1 + {\cal O}( \langle n \rangle ) ,
\label{seighteen}
\\
{\cal M}^{+}_1({\bf k}) & = & ( \varepsilon_{\bf k} - \mu ) 
+ {\cal O}( \langle n \rangle ) ,
\label{snineteen}
\\
{\cal M}^{+}_2({\bf k}) & = & ( \varepsilon_{\bf k} - \mu )^2 
\nonumber \\
& + & \Omega^2 \coth ({\frac{1}{2} \beta\Omega}) 
\sum_{\bf n} g^2({\bf n}) + {\cal O}( \langle n \rangle ).
\label{stwenty}
\end{eqnarray}
The above exact relations may be used for statistical analysis of inverse 
photoemission spectra (IPES) from low-density polaron systems.
First of all, Eq.~(\ref{seighteen}) allows proper normalization 
of the spectrum for each electron momentum {\bf k}.  According to 
Eq.~(\ref{snineteen}), the first moment yields the {\em bare}
electronic spectrum inside the crystal.  However,
in an el-ph system the chemical potential $\mu$ is lowered by
the polaron binding energy $E_p$.  As a function of ${\bf k}$, 
${\cal M}^{+}_1({\bf k})$ never gets smaller than $E_p$.  In other
words, one should expect a gap of size $E_p$ in IPES.  The second 
moment ${\cal M}^{+}_2({\bf k})$ also contains information about the 
strength of el-ph interaction.  It is convenient to parameterize
the latter in terms of the zero order (in hopping) binding energy of 
the polaron, $E^{(0)}_p \equiv \Omega \sum_{\bf n} g^2({\bf n})$, see 
\cite{AleKor}. ($E^{(0)}_p$ is employed here as a shorthand notation only.  
It does not include the effects of the kinetic energy and therefore it is 
not the true polaron energy, $E^{(0)}_p \neq E_p$.)  The mean squared 
deviation of the spectrum,
${\overline{(\Delta \omega)^2}} = {\cal M}^{+}_2 - ({\cal M}^{+}_1)^2 = 
E^{(0)}_p \Omega \coth({\frac{1}{2}\beta\Omega})$, is a direct measure
of $E^{(0)}_p$.  It has to scale approximately linearly with the IPES gap.  
Notice, that ${\overline{(\Delta \omega)^2}} = E^{(0)}_p \Omega$ at low 
temperatures and ${\overline{(\Delta \omega)^2}} \approx 2 E^{(0)}_p T$ 
at high temperatures $T$. 

One should mention that sum rules on the function $A_{-}({\bf k},\omega)$ 
have also been discussed in relation to direct photoemission. 
In particular, Randeria et al introduced an approximate sum rule for 
the zeroth moment, ${\cal M}^{-}_0 = n({\bf k})$, where $n({\bf k})$ is 
the mean occupation of the state ${\bf k}$ \cite{Randeria}.
This relation may not be generalizable to polaronic systems 
because in general $n({\bf k})$ depends on the strength
of the el-ph interaction and therefore is not known a priori. 
 
Calculation of moments can be continued.  The low density
ensures that any fermionic operator that will arise after the multiple 
commutation of $c_{\bf k}$ with the Hamiltonian, will factor out.  
The remaining phonon operators may be complex but they will have to 
be averaged over the system of free phonons.  Such an averaging can 
always be performed.  The conclusion is that {\em any} moment of 
$A_{+}({\bf k},\omega)$ can in principle be calculated analytically.  
However, the algebra quickly becomes cumbersome.  Therefore, I present
only two more moments for the Holstein model with nearest neighbor
hopping, i.e. for $g({\bf n-m}) = g_0 \delta_{\bf nm}$ and 
$\varepsilon_{\bf k} = - J \sum_{\bf l} e^{i{\bf kl}}$,
{\bf l} numbering the nearest neighbors:   
\begin{equation}
{\cal M}^{+}_3({\bf k}) = ( \varepsilon_{\bf k} - \mu )^3 +
2 g^2_T \Omega^2 ( \varepsilon_{\bf k} - \mu ) + g^2_0 \Omega^3 
+ {\cal O}( \langle n \rangle ) ,
\label{stwentyone}
\end{equation}
\begin{eqnarray}
{\cal M}^{+}_4({\bf k}) & = & ( \varepsilon_{\bf k} - \mu )^4 +  
g^2_T \Omega^2 [ 3 ( \varepsilon_{\bf k} - \mu )^2 + zJ^2 ]  \nonumber \\ 
& + & 2 g^2_0 \Omega^3 ( \varepsilon_{\bf k} - \mu ) + 
[ g^2_T + 3 g^4_T ] \Omega^4 + {\cal O}( \langle n \rangle ) ,
\label{stwentytwo}
\end{eqnarray}
where $g^2_T \equiv g^2_0 \coth({\frac{1}{2}\beta\Omega})$, and $z$ is 
the number of nearest neighbors in the lattice.  In deriving 
${\cal M}^{+}_3({\bf k})$ and ${\cal M}^{+}_4({\bf k})$ the following 
properties of the harmonic oscillator have been used:
$\langle x_{\bf m} \partial_{\bf m} \rangle = - \frac{1}{2}$, and 
$\langle x^4_{\bf m} \rangle = \frac{3}{4} \coth^2 (\frac{1}{2}\beta\Omega)$ . 
It is interesting that in the fourth moment the hopping integral 
$J$ appears not only via the bare spectrum $\varepsilon_{\bf k}$ but 
also by itself.  Thus ${\cal M}^{+}_4({\bf k})$ is the first moment 
that explicitly distinguishes the dimensionality of the lattice through 
the number of nearest neighbors $z$.  Apparently the higher order 
moments also distinguish different lattice topologies.

Apart from their relation to photoemission spectroscopy the sum 
rules may find more ``theoretical" applications.  First of all, 
they are valid for the single polaron which continues to draw 
theoretical interest.  Formal transition in the above expressions 
to the canonical ensemble with one particle is done by 
setting $\mu = 0$.  Thus the sum rules could be useful 
checks for polaron spectral functions computed numerically by exact 
diagonalization \cite{Fehske}, density matrix renormalization group
\cite{White}, quantum Monte Carlo \cite{Mishchenko}, or any other
method.  Secondly, the sum rules may provide information on the dynamics 
of polaron formation.  For instance, the quantity 
$P_{+}({\bf k},t) = |K_{+}({\bf k},t)|^2$ is the probability that the 
phonon system remains unchanged after time $t$ under the disturbance 
of an electron with momentum ${\bf k}$.  On the other hand, 
$P_{+}({\bf k},t) = 1 - 
[{\cal M}^{+}_2({\bf k})-{\cal M}^{+2}_1({\bf k})] t^2 + o(t^2) = 
1 - E^{(0)}_p \Omega \coth({\frac{1}{2}\beta\Omega}) t^2 + o(t^2)$.  
It enables one to interpret 
$\tau = [E^{(0)}_p \Omega \coth({\frac{1}{2}\beta\Omega})]^{-1/2}$ 
as the average time of phonon emission.  This is a non-perturbative 
result that is valid for any model parameters.  Finally, 
the sum rules may be employed to improve approximate analytic 
expressions for polaron spectral functions.  Consider the two 
functions ($\mu = 0$, $T = 0$):
\begin{eqnarray}
A^{\rm low}_{+}({\bf k},\omega) = (2 \pi) 
e^{-g^2_0} \sum^{\infty}_{m=0} \frac{g^{2m}_0}{m!} \times \nonumber \\
\delta(\omega + g^2_0 \Omega - e^{-g^2_0} \varepsilon_{\bf k} - m\Omega),
\label{stwentythree}
\end{eqnarray}
\begin{eqnarray}
A^{\rm high}_{+}({\bf k},\omega) = (2 \pi) 
e^{-g^2_0} \sum^{\infty}_{m=0} \frac{g^{2m}_0}{m!} \times \nonumber \\
\delta(\omega + g^2_0 \Omega - \varepsilon_{\bf k} - m\Omega) . 
\label{stwentyfour}
\end{eqnarray}
The first function is the one-polaron limit of Alexandrov and
Ranninger's result for the Holstein model in the Lang-Firsov 
approximation \cite{AleRan}.  In Eq.(\ref{stwentythree}), 
$g^2_0 \Omega$ is the zero-order binding energy $E^{(0)}_p$.  
$A^{\rm low}_{+}$ correctly describes the low-energy 
physics of the polaron, most notably the polaron band narrowing.  
However, it satisfies neither of the sum rules 
except the zeroth one.  In particular, the first moment is equal 
to $e^{-g^2_0} \varepsilon_{\bf k}$ instead of the correct 
$\varepsilon_{\bf k}$.  The second function $A^{\rm high}_{+}$
is a ``wild guess''.  It {\em does} satisfy the first and second 
moments (but not the third and fourth ones) but does not 
describe the band narrowing.  $A^{\rm high}_{+}$ is wrong at low 
energies but may be accurate at high energies.  This conjecture 
is supported by the anticipation that at high energies the electron 
should become ``free'' from phonons.  Thus a good spectral function 
could be the one that interpolates from $A^{\rm low}_{+}$ at low 
energies to $A^{\rm high}_{+}$ at high energies.  With reasonable 
parameterization, the interpolation parameters could also be deduced 
from the sum rules. 

In conclusion, a series of sum rules on the polaron spectral function
have been derived.  The sum rules are valid in the low-density limit. 
The first moment has been generalized to finite density.  A connection 
with the angle-resolved inverse photoemission spectroscopy has been 
established.  The spectrum should display a gap which size correlates 
with the mean square deviation of the spectral function.  Several other 
applications have been identified.  The sum rules are open to further 
generalizations to dispersive phonons and more complex interactions.  

I am grateful to A.\,M.\,Bratkovsky, D.\,Stewart, J.\,E.\,Hirsch, 
and especially to  A.\,S.\,Alexandrov for very useful discussions 
on the subject and critical reading of the manuscript.


\begin{references}

\bibitem{Fano}
U.\,Fano and J.\,W.\,Cooper,
Rev.\,Mod.\,Phys. {\bf 40}, 441 (1968).

\bibitem{Thole}
B.\,T.\,Thole and G.\,van der Laan,
Phys.\,Rev.\,Lett. {\bf 70}, 2499 (1993).

\bibitem{Lanzara}
A.\,Lanzara {\em et al},
Nature, {\bf 412}, 510 (2001).

\bibitem{Dessau}
Y.-D.\,Chuang {\em et al},
Science, {\bf 292}, 1509 (2001).

\bibitem{Kaminski}
A.\,Kaminski {\em et al},
Phys.\,Rev.\,Lett. {\bf 86}, 1070 (2001).

\bibitem{Kornilovitch}
A.\,S.\,Alexandrov and P.\,E.\,Kornilovitch,
Phys.\,Rev.\,Lett. {\bf 82}, 807 (1999).

\bibitem{AleKor}
A.\,S.\,Alexandrov and P.\,E.\,Kornilovitch,
cond-mat/0111549.

\bibitem{Holstein}
T.\,Holstein, 
Ann.\,Phys. {\bf 8}, 325 (1959).

\bibitem{Joynt}
R.\,Joynt,
Science, {\bf 284}, 777 (1999).

\bibitem{Mahan} 
G.\,D.\,Mahan, 
{\em Many-Particle Physics} (Plenum, New York, 1990) , 
chapter 4.

\bibitem{AleRan}
A.\,S.\,Alexandrov and J.\,Ranninger,
Phys.\,Rev.\,B {\bf 45}, 13109 (1992).

\bibitem{Sricheewin}
A.\,S.\,Alexandrov and C.\,Sricheewin,
Europhys.\,Lett. {\bf 51}, 188 (2000).

\bibitem{Perebeinos}
V.\,Perebeinos and P.\,B.\,Allen,
Phys.\,Rev.\,Lett. {\bf 85}, 5178 (2000).

\bibitem{Nolting}
W.\,Nolting,
Z.\,Phys. {\bf 255}, 25 (1972). 

\bibitem{Randeria}
M.\,Randeria {\em et al},
Phys.\,Rev.\,Lett. {\bf 74}, 4951 (1995).

\bibitem{Fehske}
G.\,Wellein and H.\,Fehske,
Phys.\,Rev.\,B {\bf 56}, 4513 (1997);
G.\,Wellein, H.\,R\"oder, and H.\,Fehske,
Phys.\,Rev.\,B {\bf 53}, 9666 (1995).

\bibitem{White}
C.\,Zhang, E.\,Jeckelmann, and S.\,R.\,White,
Phys.\,Rev.\,B {\bf 60}, 14092 (1999).

\bibitem{Mishchenko}
A.\,S.\,Mishchenko {\em et al},
Phys.\,Rev.\,B {\bf 62}, 6317 (2000).

\end{references}
\end{document}